\pdfoutput=1
\documentclass[runningheads]{llncs}
\usepackage{graphicx}
\usepackage{paperstyle}

\begin{document}
	\title{The Probabilistic Termination Tool Amber \thanks{
			This research was supported by the WWTF ICT19-018 grant ProbInG, the ERC Starting Grant
			SYMCAR 639270, ERC Consolidator Grant ARTIST 101002685, the ERC AdG Grant FRAPPANT 787914, the Austrian FWF project W1255-N23, and the SecInt Doctoral College funded by TU Wien.}}
	\author{
		Marcel Moosbrugger\inst{1}\textsuperscript{(\Letter)}\orcidID{0000-0002-2006-3741}\and
		Ezio Bartocci\inst{1}\orcidID{0000-0002-8004-6601} \and \\
		Joost-Pieter Katoen\inst{2}\orcidID{0000-0002-6143-1926} \and
		Laura Kovács\inst{1}\orcidID{0000-0002-8299-2714}
	}
	\authorrunning{M. Moosbrugger et al.}
	%
	\institute{
		TU Wien, Vienna, Austria\\
		\email{marcel.moosbrugger@tuwien.ac.at}
		\and
		RWTH Aachen University, Aachen, Germany
	}
	\maketitle              
	\begin{abstract}
		We describe  the \amber{} tool for  proving and refuting the termination of a class of probabilistic while-programs with polynomial arithmetic,  in a fully automated manner.
\amber{} combines martingale theory with properties of asymptotic bounding functions and implements relaxed versions of existing probabilistic termination proof rules to prove/disprove (positive) almost sure termination of probabilistic loops. 
\amber{} supports programs parameterized by symbolic constants and drawing from common probability distributions.
Our experimental comparisons give practical evidence of  \amber{} outperforming existing state-of-the-art tools.

	\end{abstract}

	\section{Introduction}
Probabilistic programming obviates the need to manually provide inference methods and enables rapid prototyping~\cite{ghahramani2015}.
Automated formal verification of probabilistic programs, however, is still in its infancy.
Our tool \amber{} provides a step towards solving this problem when it comes to automating the termination analysis of probabilistic programs, which is an active research topic \cite{esparzaProving2012,chakarov2013probabilistic,ferrerFioriti2015,chatterjee2016,agrawal2017lexicographic,chatterjee2017,mciver2017,huangNew2018,ChenH20}.
Probabilistic programs are almost-surely terminating (AST) if they terminate with probability $1$ on all inputs.
They are positively AST (PAST) if their expected runtime is finite \cite{bournez2005}. 
We describe \amber{}, a fully automated software artifact to prove/disprove (P)AST.
Proving (P)AST is a notoriously difficult problem; in
fact it is harder than proving traditional program termination \cite{kaminskiHardness2015}.
\amber{} supports the analysis of a class of polynomial probabilistic programs.
Programs in the supported class consist of single loops whose body is a sequence of random assignments with acyclic variable dependencies.
Moreover, \amber{}'s programming model supports programs parametrized by symbolic constants and drawing from common probability distributions.
To automate termination analysis, \amber{} automates relaxations of various existing martingale-based proof rules ensuring \mbox{(non-)}(P)AST \cite{bartheKatoenSilva2020} and combines symbolic computation with asymptotic bounding functions.
\amber{} certifies (non-)(P)AST without relying on user-provided templates/bounds over termination conditions.
Our experiments demonstrate \amber{} outperforming the state-of-the-art in the automated termination analysis of probabilistic programs (Section~\ref{sec:eval}). \\

\begin{figure}[tb]
	{\scriptsize
		$bop \in \{ +,-,*, **, / \}$,
		$cop \in \{ >, < \}$
		
		$dist \in \{$uniform, gauss, laplace, bernoulli, binomial, geometric, hypergeometric, exponential, beta, chi-squared, rayleigh$\}$
		\begin{grammar}
			<program> ::= <i_assign>$^*$ while <poly> <cop> <poly>: <rv_assign>$^+$ <v_assign>$^+$
			
			<i_assign> ::= <var> = <const> | <var> = <rv_expr> \hspace{1em} <rv_assign> ::= <var> = <rv_expr>
			
			<v_assign> ::= <var> = <branches> \hspace{1em} <rv_expr> ::= RV(<dist> [, <const>]$^*$)
			
			<branches> ::= <poly> | <poly> @ <const>; <branches>
			
			<poly> ::= $p \in C[V]$ \hspace{1em}  <sym> ::= [a-zA-Z][a-zA-Z0-9]$^*$ \hspace{1em}  <var> V ::= [a-zA-Z][a-zA-Z0-9]$^*$
			
			<const> C ::= $n \in \N$ | <sym> | - <const> | <const> <bop> <const>
		\end{grammar}
	}
	
	\caption{
		The \amber{} input syntax. 
		$C[V]$ denotes the set of polynomials in $V$ (program variables) with coefficients from $C$ (constants).
		The power operator is \lq**\rq. 
	}
	\label{fig:syntax}
\end{figure}

\noindent{\bf Related work.}
The tools \texttt{MGen}~\cite{chakarov2013probabilistic} and \texttt{LexRSM}~\cite{agrawal2017lexicographic} use linear programming techniques to certify PAST and AST, respectively.
The recent tools \texttt{Absynth}~\cite{ngo2018bounded}, \texttt{KoAT2}~\cite{meyer2021inferring} and \texttt{ecoimp}~\cite{avanzini2020modular} can establish upper bounds on expected costs, therefore also on expected runtimes, and thus certify PAST.
While powerful on respective AST/PAST domains, we note that none of the aforementioned tools support both proving and disproving (P)AST.
\amber{} is the first tool able to prove and disprove (P)AST.
Our recent work introduces relaxations of existing proof rules for probabilistic (non-)termination together with automation techniques based on \emph{asymptotic bounding functions}~\cite{moosbrugger2020automated}.
We utilize these proof rule relaxations in \amber{} and extend the technique of asymptotic bounding functions to programs
drawing from common probability distributions and including symbolic constants.\\

\noindent{\bf Contributions.}
This tool demonstration paper describes \emph{what} \amber{} can do and \emph{how} it can be used for certifying (non-)(P)AST.
\begin{itemize}
	\item We present \amber{}, a fully automatic open-source software artifact\footnote{https://github.com/probing-lab/amber} for certifying probabilistic (non-)termination (Section~\ref{sec:usage-components}).
	
	
	\item We exhaustively compare \amber{} to related tools and report on our experimental findings (Section~\ref{sec:eval}).
	
	\item We provide a benchmark suite of 50 probabilistic programs as a publicly available repository of probabilistic program examples (Section~\ref{sec:eval}).
\end{itemize}
	
	\section{Usage and Components}
\label{sec:usage-components}

\noindent{\bf Programming Model.}
\amber{} supports analyzing the probabilistic termination behavior of a class of probabilistic programs involving polynomial arithmetic and drawing from common probability distributions, parameterized by symbolic constants which represent arbitrary real numbers.
All symbolic constants are assumed to be positive.
Negative constants can be modeled with the explicit use of ``-\rq{}\rq{}.
The grammar in Figure~\ref{fig:syntax} defines the input programs to \amber{}.
Inputs consist of an initialization part and a while-loop, whose guard is a polynomial inequality over program variables.
The initialization part is a sequence of assignments either assigning (symbolic) constants or values drawn from probability distributions.
Within the loop body, program variables are updated with either (i) a value drawn from a distribution or (ii) one of multiple polynomials over program variables with some probability.
Additional to the structure imposed by the grammar in Figure~\ref{fig:syntax}, input programs are required to satisfy the following \emph{\mbox{structural constraint:}
each variable updated in the loop body depends at most linearly on itself and at most polynomially on variables preceding.}
On a high-level, this constraint enables the use of algebraic recurrence techniques for probabilistic termination analysis~\cite{moosbrugger2020automated}.
Despite the syntactical restrictions, most existing benchmarks on automated probabilistic termination analysis \cite{moosbrugger2020automated} and dynamic Bayesian networks \cite{bartocci2020analysis} can be encoded in our programming language.
Figure~\ref{fig:motivating} shows three example programs for which \amber{} is able to automatically infer the respective termination behavior.

\begin{figure}[tb]
	\begin{minipage}{0.6\linewidth}
		\begin{subfigure}{\linewidth}
			\hspace{2em}
			\begin{minipage}{\linewidth}
				\begin{Verbatim}[gobble=4, numbers=left, commandchars=\\\{\}]
				\textcolor{purple}{x} = RV(\textit{gauss}, 0, 1)
				\textcolor{purple}{y} = RV(\textit{gauss}, 0, 1)
				\textcolor{darkred}{while} x**2+y**2 < \textbf{c}:
				    \textcolor{purple}{s} = RV(\textit{uniform}, 1, 2)
				    \textcolor{purple}{t} = RV(\textit{gauss}, 0, 1)
				    \textcolor{purple}{x} = \textcolor{purple}{x}+s \textcolor{darkblue}{@1/2;} \textcolor{purple}{x}+2*s
				    \textcolor{purple}{y} = \textcolor{purple}{y}+x+t**2 \textcolor{darkblue}{@1/2;} \textcolor{purple}{y}-x-t**2
				\end{Verbatim}
			\end{minipage}
			\caption{}
			\label{fig:mot:past}
		\end{subfigure}
	\end{minipage}
	\begin{minipage}{0.4\linewidth}
		\begin{subfigure}{\linewidth}
			\begin{minipage}{\linewidth}
				\begin{Verbatim}[gobble=4, numbers=left, commandchars=\\\{\}]
				\textcolor{purple}{x} = \textbf{x0}
				\textcolor{darkred}{while} x > 0:
				    \textcolor{purple}{x} = \textcolor{purple}{x}+\textbf{c} \textcolor{darkblue}{@1/2;} \textcolor{purple}{x}-\textbf{c}
				\end{Verbatim}
			\end{minipage}
			\caption{}
			\label{fig:mot:ast}
		\end{subfigure}
		\begin{subfigure}{\linewidth}
			\vspace{1em}
			\begin{minipage}{\linewidth}
				\begin{Verbatim}[gobble=4, numbers=left, commandchars=\\\{\}]
				\textcolor{purple}{x} = \textbf{x0}
				\textcolor{darkred}{while} x > 0:
				    \textcolor{purple}{x} = \textcolor{purple}{x}+\textbf{c} \textcolor{darkblue}{@1/2+\textbf{e};} \textcolor{purple}{x}-\textbf{c}
				\end{Verbatim}
			\end{minipage}
			\caption{}
			\label{fig:mot:nast}
		\end{subfigure}
	\end{minipage}
	\caption{
		Two programs supported by \amber{}, with symbolic constants  $c,x0,e\in\R^+$;
		Program \ref{fig:mot:past} is PAST, program \ref{fig:mot:ast} is AST but not PAST and program \ref{fig:mot:nast} is not AST. 
	}
	\label{fig:motivating}
\end{figure}
\begin{figure}
	\centering
	\includegraphics[width=0.9\textwidth]{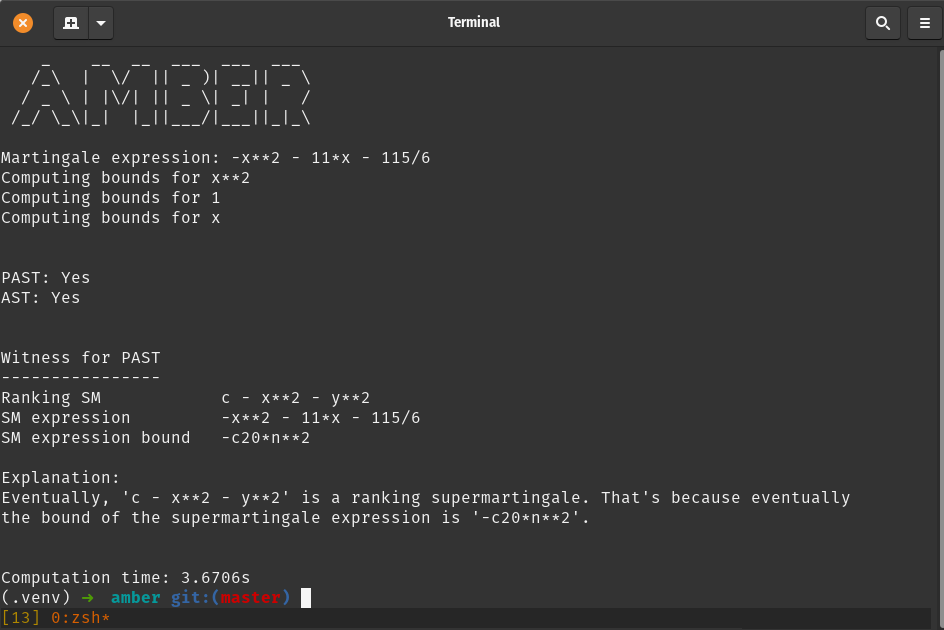}
	\caption{The output of \amber{} when run on the program from Figure~\ref{fig:mot:past}.}
	\label{fig:output}
\end{figure}


\ \\
\noindent{\bf Implementation and Usage.}
\amber{} is implemented in \texttt{python3} and relies on the \texttt{lark-parser}\footnote{\url{https://github.com/lark-parser/lark}} package to parse its input programs. Further, \amber{} uses the \texttt{diofant}\footnote{\url{https://github.com/diofant/diofant}} package as its computer-algebra system.
To compute closed-form expressions for statistical moments of monomials over program variables only depending on the loop counter, \amber{} uses the tool \texttt{Mora}~\cite{BartocciKS20}.
However, for efficient integration within \amber{}, we reimplemented and adapted the \texttt{Mora} functionalities exploited by \amber{} (\texttt{Mora v2}), in particular by employing dynamic programming to avoid redundant computations.
Altogether, \amber{} consists of $\sim 2000$ lines of code.
Figure~\ref{fig:output} shows \amber{}'s output when run on the program from Figure~\ref{fig:mot:past}.
\amber{} can be used through a Docker container~\cite{merkel2014docker} or installed locally.
Detailed installation and usage instructions are available at \url{https://github.com/probing-lab/amber}.

\ \\
\noindent{\bf Run with Docker.}
\amber{} can be used through a Docker container~\cite{merkel2014docker} by running:
\texttt{\$ docker run -ti marcelmoosbrugger/amber}

\noindent
\amber{} can be run on our \texttt{2d\_bounded\_random\_walk} benchmark with:
\ \\
\texttt{\$ ./amber benchmarks/past/2d\_bounded\_random\_walk}

\ \\
\noindent{\bf Local installation.}
First, clone the repository by running the following command in your terminal:
\texttt{\$ git clone git@github.com:probing-lab/amber.git}

\noindent
Change directories to \amber{}'s root folder and make sure \texttt{python3.8} and the package manager \texttt{pip} are installed on your system.
All required python packages can be installed by running
\texttt{\$ pip install -r requirements.txt}

\noindent
Create an input program (see Section~\ref{sec:usage-components}) and save it in the \texttt{benchmarks} folder for example with the file name \texttt{my-benchmark}.
\amber{} can now be run with respect to the input program \texttt{benchmarks/my-benchmark} with the following command:
\texttt{\$ python ./amber.py -{}-benchmarks benchmarks/my-benchmark}

\begin{figure}
	\centering
	\includegraphics[width=\textwidth,height=\textheight,keepaspectratio]{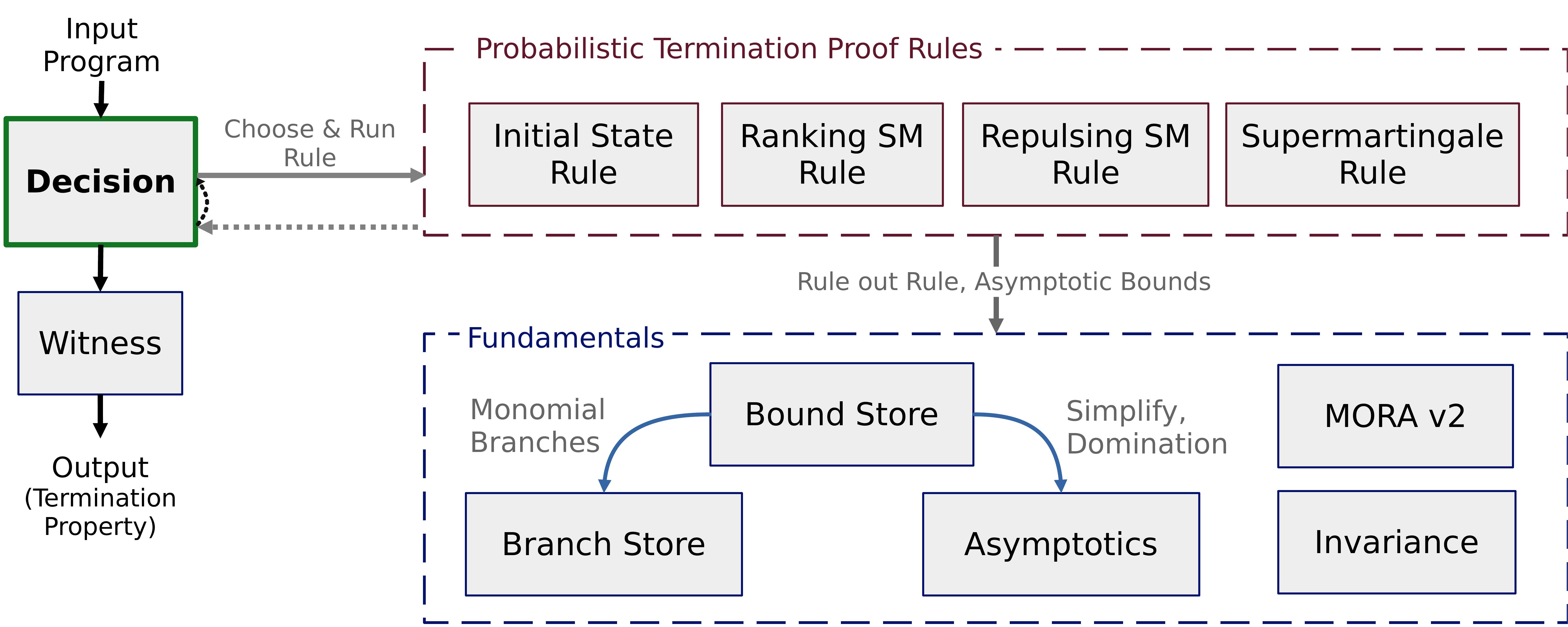}
	\caption{Main components of \amber{} and interactions between them.}
	\label{fig:components}
\end{figure}

\ \\
\noindent{\bf Components.}
Figure~\ref{fig:components} illustrates \amber{}'s main components.
\amber{} uses four existing probabilistic termination proof rules~\cite{chakarov2013probabilistic,ferrerFioriti2015,mciver2017,chatterjee2017} and their relaxations~\cite{moosbrugger2020automated}.
Additionally, \amber{} extends the algorithms for these relaxations to further support drawing from common probability distributions and symbolic constants (cf. Figure~\ref{fig:syntax}).
After parsing the input program, \amber{} initializes the four proof rule relaxations and determines their applicability~\cite{moosbrugger2020automated}.
\amber{} then executes applicable proof rules consecutively and reports the analysis result containing potential witnesses for \mbox{(non-)(P)AST}.
The proof rule algorithms require the computation of asymptotic bounding functions which is implemented in the \emph{Bound Store} component.

	\section{Evaluation}
\label{sec:eval}

\begin{table}[t]
	\tablesize
	\centering
	\bgroup
	\def\arraystretch{1.5}
	\begin{tabular}{lcccccc}
		\toprule
		Program & \rotatebox{\tablerotation}{\amber{}} & \rotatebox{\tablerotation}{\texttt{Absynth}} & \rotatebox{\tablerotation}{\texttt{MGen}} & \rotatebox{\tablerotation}{\texttt{LexRSM}} & \rotatebox{\tablerotation}{\texttt{KoAT2}} & \rotatebox{\tablerotation}{\texttt{ecoimp}} \\
		\midrule
		2d\_bounded\_random\_walk & \benchsucc{} & \benchfail{} & \benchna{} & \benchna{} & \benchfail{} & \benchfail{} \\
		\hdashline
		biased\_random\_walk\_const & \benchsucc{} & \benchsucc{} & \benchsucc{} & \benchsucc{} & \benchsucc{} & \benchsucc{} \\
		\hdashline
		biased\_random\_walk\_exp & \benchsucc{} & \benchfail{} & \benchsucc & \benchfail{} & \benchfail{} & \benchfail{} \\
		\hdashline
		biased\_random\_walk\_poly & \benchsucc{} & \benchfail{} & \benchfail & \benchna{} & \benchfail{} & \benchfail{} \\
		\hdashline
		binomial\_past & \benchsucc{} & \benchsucc{} & \benchsucc{} & \benchsucc{} & \benchsucc{} & \benchsucc{} \\
		\hdashline
		complex\_past & \benchsucc{} & \benchfail{} & \benchna{} & \benchna{} & \benchfail{} & \benchfail{} \\
		\hdashline
		consecutive\_bernoulli\_trails & \benchsucc{} & \benchsucc{} & \benchsucc{} & \benchsucc{} & \benchsucc{} & \benchsucc{} \\
		\hdashline
		coupon\_collector\_4 & \benchsucc{} & \benchfail{} & \benchsucc{} & \benchsucc{} & \benchsucc{} & \benchsucc{} \\
		\hdashline
		coupon\_collector\_5 & \benchsucc{} & \benchfail{} & \benchsucc{} & \benchsucc{} & \benchsucc{} & \benchsucc{} \\
		\hdashline
		dueling\_cowboys & \benchsucc{} & \benchsucc{} & \benchsucc{} & \benchsucc{} & \benchsucc{} & \benchsucc{} \\
		\hdashline
		exponential\_past\_1 & \benchsucc{} & \benchna{} & \benchna{} & \benchna{} & \benchfail{} & \benchna{} \\
		\hdashline
		exponential\_past\_2 & \benchsucc{} & \benchna{} & \benchna{} & \benchna{} & \benchfail{} & \benchna{} \\
		\hdashline
		geometric & \benchsucc{} & \benchsucc{} & \benchsucc{} & \benchsucc{} & \benchsucc{} & \benchsucc{} \\
		\hdashline
		geometric\_exp & \benchfail{} & \benchfail{} & \benchfail{} & \benchfail{} & \benchfail{} & \benchfail{} \\
		\hdashline
	\end{tabular}\hfill
	\begin{tabular}{lcccccc}
		\midrule
		Program & \rotatebox{\tablerotation}{\amber{}} & \rotatebox{\tablerotation}{\texttt{Absynth}} & \rotatebox{\tablerotation}{\texttt{MGen}} & \rotatebox{\tablerotation}{\texttt{LexRSM}} & \rotatebox{\tablerotation}{\texttt{KoAT2}} & \rotatebox{\tablerotation}{\texttt{ecoimp}} \\
		\midrule
		linear\_past\_1 & \benchsucc{} & \benchfail{} & \benchfail{} & \benchfail{} & \benchfail{} & \benchfail{} \\
		\hdashline
		linear\_past\_2 & \benchsucc{} & \benchfail{} & \benchna{} & \benchfail{} & \benchfail{} & \benchfail{} \\
		\hdashline
		nested\_loops & \benchna{} & \benchsucc & \benchfail{} & \benchsucc{} & \benchsucc{} & \benchsucc{} \\
		\hdashline
		polynomial\_past\_1 & \benchsucc{} & \benchfail{} & \benchna{} & \benchna{} & \benchfail{} & \benchfail{} \\
		\hdashline
		polynomial\_past\_2 & \benchsucc{} & \benchfail{} & \benchna{} & \benchna{} & \benchfail{} & \benchfail{} \\
		\hdashline
		sequential\_loops & \benchna{} & \benchsucc{} & \benchfail{} & \benchsucc{} & \benchsucc{} & \benchsucc{} \\
		\hdashline
		tortoise\_hare\_race & \benchsucc{} & \benchsucc{} & \benchsucc & \benchsucc{} & \benchsucc{} & \benchsucc{} \\
		\midrule
		dependent\_dist* & \benchna{} & \benchna{} & \benchna{} & \benchna{} & \benchfail{} & \benchsucc{} \\
		\hdashline
		exp\_rw\_gauss\_noise* & \benchsucc{} & \benchna{} & \benchna{} & \benchna{} & \benchna{} & \benchna{} \\
		\hdashline
		gemoetric\_gaussian* & \benchsucc{} & \benchna{} & \benchna{} & \benchna{} & \benchna{} & \benchna{} \\
		\hdashline
		race\_uniform\_noise* & \benchsucc{} & \benchfail{} & \benchsucc{} & \benchsucc{} & \benchfail{} & \benchsucc{} \\
		\hdashline
		symb\_2d\_rw* & \benchsucc{} & \benchfail{} & \benchna{} & \benchna{} & \benchfail{} & \benchfail{} \\
		\hdashline
		uniform\_rw\_walk* & \benchsucc{} & \benchsucc{} & \benchsucc{} & \benchsucc{} & \benchsucc{} & \benchsucc{} \\
		\midrule
		Total \benchsucc{} & 23 & 9 & 11 & 12 & 11 & 13 \\
		\bottomrule
	\end{tabular}
	\egroup
	\vspace{1em}
	\caption{27 programs which are PAST.}
	\label{tab:bench:past}
\end{table}

\begin{figure}[t!]
	\begin{minipage}{0.45\linewidth}
		\begin{table}[H]
			\tablesize
			\centering
			\bgroup
			\def\arraystretch{1.5}
			\begin{tabular}{lcc}
				\toprule
				Program & \rotatebox{\tablerotation}{\amber{}} & \rotatebox{\tablerotation}{\texttt{LexRSM}} \\
				\midrule
				fair\_in\_limit\_random\_walk & \benchna{} & \benchna{} \\
				\hdashline
				gambling & \benchsucc{} & \benchfail{} \\
				\hdashline
				symmetric\_2d\_random\_walk & \benchfail{} & \benchna{} \\
				\hdashline
				symmetric\_random\_walk\_constant\_1 & \benchsucc{} & \benchfail{} \\
				\hdashline
				symmetric\_random\_walk\_constant\_2 & \benchsucc{} & \benchfail{} \\
				\hdashline
				symmetric\_random\_walk\_exp\_1 & \benchsucc{} & \benchfail{} \\
				\hdashline
				symmetric\_random\_walk\_exp\_2 & \benchsucc{} & \benchna{} \\
				\hdashline
				symmetric\_random\_walk\_linear\_1 & \benchsucc{} & \benchfail{} \\
				\hdashline
				symmetric\_random\_walk\_linear\_2 & \benchsucc{} & \benchfail{} \\
				\hdashline
				symmetric\_random\_walk\_poly\_1 & \benchsucc{} & \benchna{} \\
				\hdashline
				symmetric\_random\_walk\_poly\_2 & \benchsucc{} & \benchna{} \\
				\midrule
				gaussian\_rw\_walk* & \benchsucc{} & \benchna{} \\
				\hdashline
				laplacian\_noise* & \benchsucc{} & \benchna{} \\
				\hdashline
				symb\_1d\_rw* & \benchsucc{} & \benchna{} \\
				\midrule
				Total \benchsucc{} & 12 & 0 \\
				\bottomrule
			\end{tabular}
			\egroup
			\vspace{1em}
			\caption{14 programs which are AST and not necessarily PAST.}
			\label{tab:bench:ast}
		\end{table}
	\end{minipage}\hfill
	\begin{minipage}{0.45\linewidth}
		\begin{table}[H]
			\tablesize
			\centering
			\bgroup
			\def\arraystretch{1.5}
			\begin{tabular}{lc}
				\toprule
				Program & \amber{} \\
				\midrule
				biased\_random\_walk\_nast\_1 & \benchsucc{} \\
				\hdashline
				biased\_random\_walk\_nast\_2 & \benchsucc{} \\	
				\hdashline
				biased\_random\_walk\_nast\_3 & \benchsucc{} \\
				\hdashline
				biased\_random\_walk\_nast\_4 & \benchsucc{} \\
				\hdashline
				binomial\_nast & \benchsucc{} \\
				\hdashline
				polynomial\_nast & \benchfail{} \\
				\midrule
				binomial\_nast\_noise* & \benchsucc{} \\
				\hdashline
				symb\_nast\_1d\_rw* & \benchsucc{} \\
				\hdashline
				hypergeo\_nast* & \benchsucc \\
				\midrule
				Total \benchsucc{} & 8 \\
				\bottomrule
			\end{tabular}
			\egroup
			\vspace{1em}
			\caption{9 programs which are not AST.}
			\label{tab:bench:nast}
		\end{table}
	\end{minipage}
\end{figure}

\noindent{\bf Experimental Setup.}
\amber{} and our benchmarks, are publicly available at \url{https://github.com/probing-lab/amber}.
The output of \amber{} is an answer (``Yes\rq{}\rq{}, ``No\rq{}\rq{} or ``Maybe\rq{}\rq{}) to PAST and AST, together with a potential witness.
We took all $39$ benchmarks from \cite{moosbrugger2020automated} and extended them by $11$ new programs to test \amber{}'s capability to handle symbolic constants and drawing from probability distributions.
The $11$ new benchmarks are constructed from the $39$ original programs, by adding noise drawn from common probability distributions and replacing concrete constants with symbolic ones.
As such, we conduct experiments using a total of $50$ challenging benchmarks, involving polynomial arithmetic, probability distributions and symbolic constants.
Further, we compare \amber{} not only against \texttt{Absynth} and \texttt{MGen} (as in \cite{moosbrugger2020automated}), but also evaluate \amber{} in comparison to the recent tools \texttt{LexRSM}~\cite{agrawal2017lexicographic}, \texttt{KoAT2}~\cite{meyer2021inferring} and \texttt{ecoimp}~\cite{avanzini2020modular}.
Note that \texttt{MGen} can only certify PAST and \texttt{LexRSM} only AST.
Moreover, the tools \texttt{Absynth}, \texttt{KoAT2} and
\texttt{ecoimp} mainly aim to find upper bounds on expected costs.
Tables~\ref{tab:bench:past}-\ref{tab:bench:nast} summarize our experimental results, with benchmarks separated into \emph{PAST} (Table~\ref{tab:bench:past}), \emph{AST but not PAST} (Table~\ref{tab:bench:ast}), and \emph{not AST} (Table~\ref{tab:bench:nast}).
Benchmarks marked with * are part of our $11$ new examples.
In every table, \benchsucc{} (\benchfail) marks a tool (not) being able to certify the respective termination property.
Moreover, \benchna{} symbolizes that a benchmark is out-of-scope for a tool, for instance, due to not supporting some distributions or polynomial arithmetic.
All benchmarks have been run on a machine with a 2.6 GHz Intel i7 (Gen 10) processor and 32 GB of RAM and finished within a timeout of 50 seconds, where most experiments terminated within a few seconds.\\

\noindent{\bf Experimental Analysis.}
\amber{} successfully certifies $23$ out of the $27$ PAST benchmarks (Table~\ref{tab:bench:past}).
Although \texttt{Absynth}, \texttt{KoAT2} and \texttt{ecosimp} can find expected cost upper bounds for large programs~\cite{ngo2018bounded,meyer2021inferring,avanzini2020modular}, they struggle on small programs whose termination is not known a priori.
For instance, they struggle when a benchmark probabilistically ``chooses\rq{}\rq{} between two polynomials working against each other (one moving the program state away from a termination criterion and one towards it).
Our experiments show that \amber{} handles such cases successfully.
\texttt{MGen} supports the continuous uniform distribution and \texttt{KoAT2} the geometric distribution whose support is infinite.
With these two exceptions, \amber{} is the only tool supporting continuous distributions and distributions with infinite support.
To the best of our knowledge, \amber{} is the first tool certifying PAST supporting both discrete and continuous distributions as well as distributions with finite and infinite support.
\amber{} successfully certifies $12$ benchmarks to be AST which are not PAST (Table~\ref{tab:bench:ast}).
Whereas the \texttt{LexRSM} tool can certify non-PAST programs to be AST, such programs need to contain subprograms which are PAST \cite{agrawal2017lexicographic}.
The well-known example of \texttt{symmetric\_1D\_random\_walk}, contained in our benchmarks, does not have a PAST subprogram.
Therefore, the \texttt{LexRSM} tool cannot establish AST for it.
In contrast, \amber{} using the \emph{Supermartingale Rule} can handle these programs.
To the best of our knowledge,
\amber{} is the first tool capable of certifying non-AST for polynomial probabilistic programs involving drawing from distributions and symbolic constants.
\amber{} is also the first tool automating (non-)AST and (non-)PAST analysis in a unifying manner.\\

\noindent{\bf Experimental Summary.}
Tables~\ref{tab:bench:past}-\ref{tab:bench:nast} demonstrate that (i) 
\amber{} outperforms the state-of-the-art in certifying (P)AST, and (ii) 
\amber{} determines \mbox{(non-)(P)AST} for programs with various distributions and symbolic constants.

	\section{Conclusion}
We described \amber{}, an open-source tool for analyzing the termination behavior for polynomial probabilistic programs, in a fully automatic way.
\amber{} computes asymptotic bounding functions and martingale expressions and is the first tool to prove and disprove (P)AST in a unifying manner. 
\amber{} can analyze continuous, discrete, finitely- and infinitely supported distributions in polynomial probabilistic programs parameterized by symbolic constants.
Our experimental comparisons give practical evidence that \amber{} can (dis)prove (P)AST for a substantially larger class of programs than state-of-the-art tools.

	%
	%
	%
	\bibliographystyle{splncs04}

	\bibliography{paper}
\end{document}